\documentclass[aps,prc,twocolumn,superscriptaddress,showpacs,showkeys]{revtex4}
\usepackage{graphicx,amsmath,amssymb,bm}

\newcommand{\be}{\begin{equation}}
\newcommand{\ee}{\end{equation}}

\newcommand{\mev}{\, \text{MeV}}
\newcommand{\gcq}{\, \text{g}/\text{cm}^{3}}
\newcommand{\bnt}{b_{n{^3}\hspace*{-0.2mm}{\rm{H}}}}
\newcommand{\bnhe}{b_{n{^3}\hspace*{-0.2mm}{\rm{He}}}}
\newcommand{\bpt}{b_{p{^3}\hspace*{-0.2mm}{\rm{H}}}}
\newcommand{\bphe}{b_{p{^3}\hspace*{-0.2mm}{\rm{He}}}}
\newcommand{\zt}{z_{{^3}\hspace*{-0.2mm}{\rm{H}}}}
\newcommand{\zhe}{z_{{^3}\hspace*{-0.2mm}{\rm{He}}}}
\newcommand{\lamba}{\lambda_{\alpha}}
\newcommand{\lambt}{\lambda_{{^3}\hspace*{-0.2mm}{\rm{H}}}}
\newcommand{\lambhe}{\lambda_{{^3}\hspace*{-0.2mm}{\rm{He}}}}
\newcommand{\tnu}{{T_\nu}}

\begin{document}

\title{Neutrino Breakup of $A=3$ Nuclei in Supernovae}

\author{E.~O'Connor}
\email[E-mail:~]{evanoc@triumf.ca}
\affiliation{TRIUMF, 4004 Wesbrook Mall, Vancouver, BC, Canada, V6T 2A3}
\affiliation{Department of Physics, UPEI, 550 University Ave, 
Charlottetown, PE, Canada, C1A 4P3}
\author{D.~Gazit}
\email[E-mail:~]{gdoron@phys.huji.ac.il}
\affiliation{Racah Institute of Physics, Hebrew University, 91904, 
Jerusalem, Israel}
\author{C.J.~Horowitz}
\email[E-mail:~]{horowit@indiana.edu}
\affiliation{Nuclear Theory Center and Department of Physics, 
Indiana University, Bloomington, IN 47408}
\author{A.~Schwenk}
\email[E-mail:~]{schwenk@triumf.ca}
\affiliation{TRIUMF, 4004 Wesbrook Mall, Vancouver, BC, Canada, V6T 2A3}
\author{N.~Barnea}
\email[E-mail:~]{nir@phys.huji.ac.il}
\affiliation{Racah Institute of Physics, Hebrew University, 91904, 
Jerusalem, Israel}

\begin{abstract}
We extend the virial equation of state to include $^3$H and $^3$He
nuclei, and predict significant mass-three fractions near the 
neutrinosphere in supernovae. While alpha particles are often more 
abundant, we demonstrate that energy transfer cross-sections for muon
and tau neutrinos at low densities are dominated by breakup of the 
loosely-bound $^3$H and $^3$He nuclei. The virial coefficients
involving $A=3$ nuclei are calculated directly from the corresponding
nucleon-$^3$H and nucleon-$^3$He scattering phase shifts. For the
neutral-current inelastic cross-sections and the energy transfer 
cross sections, we
perform ab-initio calculations based on microscopic two- and
three-nucleon interactions and meson-exchange currents.
\end{abstract}


\maketitle

\section{Introduction}

Core-collapse supernovae (SN) are giant explosions of massive stars 
that radiate $99\%$ of their energy in neutrinos. Therefore, the
dynamics and the neutrino signals can be sensitive to the details 
of neutrino interactions with nucleonic matter. At present, most 
SN simulations with detailed neutrino microphysics do not explode, 
but they may be close to successful explosions 
(for a status report, see~\cite{Janka1,Janka2}). Supernovae
radiate electron, muon and tau neutrinos. Electron neutrinos can 
exchange energy with matter via charged-current interactions. Energy
transfer from muon or tau neutrinos, hereafter $\nu_x$, is more 
difficult~\cite{Bruenn} because neutrino-electron scattering has a 
small cross section, and neutrino-nucleon elastic scattering involves 
only a small energy transfer.

Haxton and Bruenn proposed that $\nu_x$ can exchange energy via 
inelastic excitations of $^4$He and heavier nuclei~\cite{Haxton},
whereas Hannestad and Raffelt investigated the exchange of energy 
between $\nu_x$ and two interacting nucleons $\nu_x{\rm NN} \rightarrow
\nu_x{\rm NN}$~\cite{Raffelt}. Recently, Juodagalvis {\it et al.}
have calculated detailed $\nu_x$ neutral-current 
cross-sections for $A=50-65$
nuclei~\cite{Juodagalvis}, and Gazit and Barnea have presented
microscopic results for $^4$He cross sections~\cite{Gazit1,Gazit2}. These 
inelastic excitations can aid the transfer of neutrino energy to 
the SN shock and can keep $\nu_x$ in thermal equilibrium to lower 
densities, resulting in the radiation of a lower energy $\nu_x$ 
spectrum~\cite{Keil}. First studies on the role of $^4$He excitation
for the shock revival were carried out by Ohnishi {\it et 
al.}~\cite{Ohnishi}.

To evaluate the role of $\nu_x$ inelastic scattering one needs 
both cross sections and detailed information on the composition
and other thermodynamic properties of nucleonic matter. Models
that describe the system with only a single average nuclear species 
may miss the contribution of less abundant nuclei with large
cross sections. Moreover, there are many fundamental connections
between the equation of state and neutrino interactions.

Nuclear statistical equilibrium (NSE) models predict abundances 
based on binding energies and the quantum numbers of nuclei. However,
NSE models only treat approximately (or neglect) strong interactions
between nuclei, and consequently break down as the density increases.
We have recently developed a description of low-density nuclear matter 
(composed of neutrons, protons and alpha particles) in thermal
equilibrium based on the virial expansion~\cite{vEOSnuc,vEOSneut}.
The virial equation of state systematically takes into account 
contributions from bound nuclei and the scattering continuum, and 
thus provides a framework to include strong-interaction corrections 
to NSE models. The virial equation of state makes model-independent
predictions for the conditions~\cite{sn1987a} 
near the neutrinosphere, for low 
densities $\rho \sim 10^{11-12} \gcq$ and high temperatures $T \sim 
4 \mev$. In particular, the resulting alpha particle concentration 
differs from all equations of state currently used in SN
simulations, and the predicted large symmetry energy at low densities 
has been confirmed in near Fermi-energy heavy-ion collisions~\cite{sym}.
In addition, the long-wavelength neutrino response of low-density matter 
can be calculated consistently from the virial expansion~\cite{vResponse}.

In this paper, we extend the virial expansion to include $^3$H and 
$^3$He nuclei, and predict that the mass-three fraction can be 
significant (up to $10 \%$) near the neutrinosphere.
The second virial coefficients involving $A=3$ nuclei are calculated 
directly from the corresponding nucleon-$^3$H and nucleon-$^3$He 
scattering phase shifts. While alpha particles are often more 
abundant due to the large binding energy ($E_4 = 28.3 \mev$
compared to $E_3 \sim 8 \mev$), we show that mass-three nuclei are 
important for energy transfer, in particular for the more energetic 
muon and tau neutrinos with $E_{\nu_x} \sim 20 \mev$. For neutrinos
with these energies, we find that the neutral-current inelastic
energy transfer cross-sections and the neutrino energy loss for 
$T \gtrsim 4 \mev$ are dominated by the breakup of the 
loosely-bound $^3$H and $^3$He nuclei. Our predictions for
the neutral-current inclusive inelastic
cross-sections on mass-three nuclei are based
on microscopic two- and three-nucleon interactions and 
meson-exchange currents, and include full final-state interactions
via the Lorentz integral transform (LIT) method~\cite{LIT}.

This paper is organized as follows. In Section~\ref{comp}, we generalize
the virial equation of state to include $A=3$ nuclei, and present results 
for the composition of low-density nuclear matter for various temperatures, 
densities and proton fractions. In Section~\ref{cross}, we calculate
the inelastic $^3$H and $^3$He neutral-current cross-sections and energy
transfer cross-sections. We combine our results in Section~\ref{dEdx}
and study the neutrino energy loss for conditions near the neutrinosphere.
Finally, we conclude in Section~\ref{conclusions}.

\section{Composition of low-density nuclear matter}
\label{comp}

In this section, we discuss the virial equation of state and present
results for the composition of low-density nuclear matter including 
$A=3$ nuclei (for details and additional thermodynamic results, 
see~\cite{A3virial}).

To determine the abundance of $A=3$ nuclei near the neutrinosphere in 
supernovae, we extend the virial approach of Refs.~\cite{vEOSnuc,vEOSneut} 
to include $^3$H and $^3$He nuclei. In the corresponding 
virial expansion, neutrons, protons, $\alpha$ particles, 
$^3$H and $^3$He nuclei are explicitly included, deuterons 
are included as a bound state contribution to the proton-neutron 
virial coefficient. We will explicitly consider deuterons and 
neutrino-deuteron scattering in future work. The equation of state is 
determined through an expansion of the pressure P in the fugacities
(see for instance~\cite{Huang}) up to second order,
\begin{align}
\frac{P}{T} &= \frac{2}{\lambda^3_{\rm N}} \bigl( z_n + z_p 
+ (z_n^2+z_p^2) \, b_n + 2 z_n z_p \, b_{pn} \bigr) \nonumber \\
&+ \frac{1}{\lamba^3} \bigl( z_\alpha
+ z_\alpha^2 \, b_\alpha + 2 z_\alpha (z_n+z_p) \,
b_{\alpha n} \bigr) \nonumber \\
&+ \frac{2}{\lambhe^3} \bigl(\zhe + 2 \zhe (z_p \, \bphe + z_n \, \bnhe)
\bigr) \nonumber \\
&+ \frac{2}{\lambt^3} \bigl( \zt + 2 \zt (z_p \, \bpt + z_n \, \bnt) 
\bigr) \,,
\end{align}
where T is the temperature, $z_i = e^{(\mu_i + E_i) / T}$ is the fugacity 
(with chemical potential $\mu_i$ and binding energy $E_i$), 
$\lambda_i=\sqrt{2\pi / m_iT}$ is the 
thermal wavelength of particle $i$, and $b_{ij}$ are the second virial 
coefficients describing interactions between particles $i,j$
($b_i \equiv b_{ii}$). We have
calculated these virial coefficients from phases shifts and low-energy 
scattering lengths, see Ref.~\cite{A3virial} for details. The second virial 
coefficients $b_{{n}}$, $b_{{pn}}$, $b_{{\alpha}}$ and $b_{{\alpha n}}$ 
are tabulated in~\cite{vEOSnuc}, and the additional virial coefficients 
involving $A=3$ nuclei are given in Table~\ref{tab:A3b2} from $T = 1\mev$
to $10\mev$. Due to a lack of $p$-$^3$H scattering data, we assume that 
$\bpt \approx \bnhe$. The effects of interactions with $A=3$ 
nuclei are expected to be small at low densities (see also the hierarchy 
observed in Ref.~\cite{vEOSnuc}) 
and therefore we neglect $b_{{^3}\hspace*{-0.2mm}{\rm{He}}
{^3}\hspace*{-0.2mm}{\rm{He}}}$, $b_{{^3}\hspace*{-0.2mm}{\rm{H}}
{^3}\hspace*{-0.2mm}{\rm{H}}}$, $b_{{^3}\hspace*{-0.2mm}{\rm{He}}
{^3}\hspace*{-0.2mm}{\rm{H}}}$, $b_{\alpha{^3}\hspace*{-0.2mm}{\rm{He}}}$ 
and $b_{\alpha{^3}\hspace*{-0.2mm}{\rm{H}}}$.

\begin{table}
\begin{ruledtabular}
\begin{tabular}{c|p{1.5cm}|p{1.5cm}|p{1.5cm}}
$T\, \text{[MeV]}$ & $\bphe$ & $\bnt$ & $\bnhe$ \\[1mm] \hline
1 & 0.043 & -0.002 & -0.244 \\
2 & 0.245 & 0.253 & -0.067 \\
3 & 0.407 & 0.429 & 0.045 \\
4 & 0.528 & 0.555 & 0.113 \\
5 & 0.617 & 0.645 & 0.153 \\
6 & 0.680 & 0.708 & 0.176 \\
7 & 0.725 & 0.751 & 0.189 \\
8 & 0.754 & 0.778 & 0.195 \\
9 & 0.772 & 0.795 & 0.198 \\
10 & 0.781 & 0.803 & 0.198 \\
\end{tabular}
\end{ruledtabular}
\caption{Nucleon-$^3$H and nucleon-$^3$He virial coefficients for 
temperatures of $T = 1 \mev$ to $10 \mev$.}
\label{tab:A3b2}
\end{table}

\begin{figure*}[t]
\begin{center}
\includegraphics[scale=0.45,clip=]{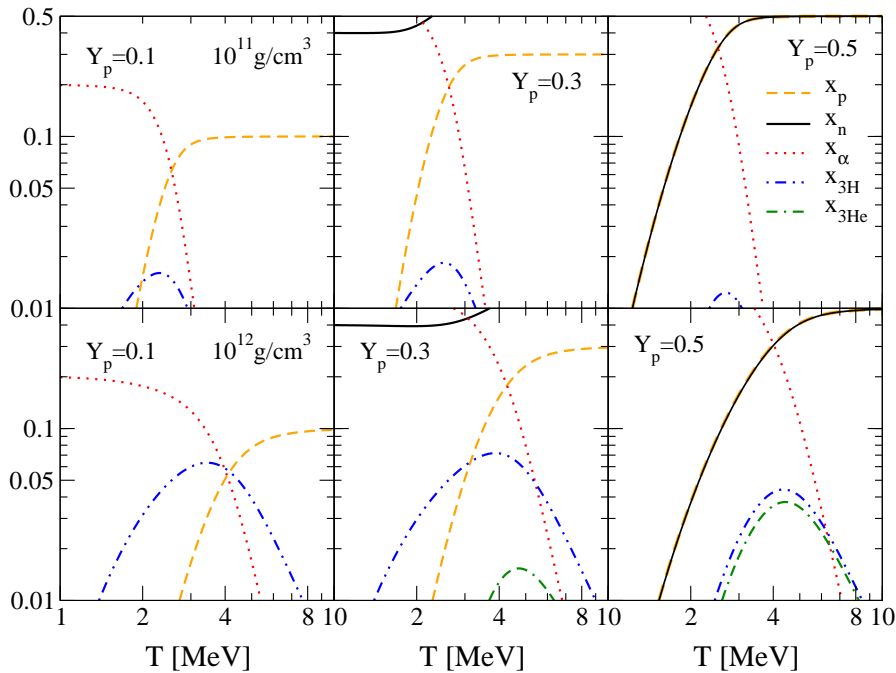}
\caption{(Color online) Mass fractions of nucleons and $A=3,4$ nuclei 
in chemical equilibrium as a function of temperature T. The top and 
bottom rows correspond to a density of $10^{11}\gcq$ and $10^{12}\gcq$ 
respectively, and from left to right the proton fractions are $Y_p = 
0.1, 0.3$ and $0.5$. For this temperature range, at  $10^{11}\gcq$ 
the neutron fugacity is $z_n < 0.1, 0.05, 0.01$ for $Y_p = 0.1, 0.3$
and $0.5$ respectively, all other fugacities are $<0.01$ for all shown 
proton fractions. At the higher density, $10^{12}\gcq$, the neutron 
fugacity is $z_n < 0.7, 0.35, 0.04$ for $Y_p = 0.1, 0.3$ and $0.5$ 
respectively, all other fugacities are $<0.06$ for all shown proton 
fractions.}
\label{fig:composition}
\end{center}
\end{figure*}

The particle densities can be obtained by partial derivatives of 
the pressure with respect to the corresponding fugacities,
\begin{equation}
n_i = z_i \left(\frac{\partial}{\partial z_i} \frac{P}{T} 
\right)_T \,.
\end{equation}
By assuming chemical equilibrium,
\begin{align}
\mu_\alpha &= 2\mu_n+2\mu_p \,, \\[2mm]
\mu_{{^3}\hspace*{-0.2mm}{\rm{H}}} &= 2\mu_n+\mu_p \,, \\[2mm]
\mu_{{^3}\hspace*{-0.2mm}{\rm{He}}} &= \mu_n+2\mu_p \,,
\end{align}
the corresponding $A=3,4$ fugacities are determined by the proton 
and neutron fugacities and the corresponding binding energy:
$\zt = z_n^2 z_p e^{E_{{^3}\hspace*{-0.2mm}{\rm{H}}}/T}$ 
for example. For a given baryon 
density $n_b$ and proton fraction $Y_p$, the fugacities $z_p$ and 
$z_n$ are then determined implicitly from
\begin{align}
n_b &= n_p + n_n+ 4n_\alpha + 3n_{{^3}\hspace*{-0.2mm}{\rm{He}}} 
+ 3n_{{^3}\hspace*{-0.2mm}{\rm{H}}} \,, \\[2mm]
Y_p &= (n_p + 2n_\alpha + 2n_{{^3}\hspace*{-0.2mm}{\rm{He}}} + 
n_{{^3}\hspace*{-0.2mm}{\rm{H}}})/n_b \,.
\end{align}
The resulting mass fractions ($x_i = A_in_i/n_b$) are shown in 
Fig.~\ref{fig:composition} for neutrinosphere densities and 
temperatures, and various proton fractions.

For a density of $10^{12}\gcq$, we find that the total mass-three
fraction is significant, up to $x_{{^3}\hspace*{-0.2mm}{\rm{H}}} 
+ x_{{^3}\hspace*{-0.2mm}{\rm{He}}} \approx 
0.1$ in symmetric matter. Moreover, for temperatures $T \gtrsim 5
\mev$ (and increasingly important at low proton fractions), the $^3$H
mass fraction is larger than the $\alpha$ particle fraction.
We also observe that at lower temperatures, where many of the protons
are bound in alpha particles (and heavy nuclei), $^3$H nuclei are more 
abundant than free protons. Here, the rate of electron capture may 
be dominated by capture on $^3$H, since the capture on $^4$He has
low cross sections and there are few free protons. The contribution
of mass-three nuclei to charged-current interactions will be left to 
future work~\cite{charged}. Finally, we emphasize that heavier nuclei
and larger clusters become important as the $\alpha$ particle
fraction saturates at very low temperature.

\section{Neutrino-$^3{\rm{\bf H}}$ and -$^3{\rm{\bf He}}$ breakup:
neutral-current cross-sections}

The calculation of the neutral-current inclusive inelastic
cross-sections on $^3$H and 
$^3$He follows Ref.~\cite{Gazit2}. We solve the three-nucleon
problem based on the Argonne $v_{18}$ nucleon-nucleon~\cite{AV18}
and the Urbana IX three-nucleon~\cite{UIX} interactions. Neutrino
scattering on $A=3$ nuclei only induces transitions to continuum 
states, since $^3$H and $^3$He have no excited states. Hence,
a correct description must include breakup channels and 
final-state interactions among the three nucleons. We include these
via the LIT method~\cite{LIT}, which uses an integral transform 
with a lorentzian kernel to reduce the continuum problem to a
bound-state-like problem. The resulting Schr\"{o}dinger-like 
equations are solved using the effective interaction hyperspherical 
harmonics (EIHH) approach~\cite{EIHH,symhh}. The combination of 
these approaches converges rapidly and yields a numerical precision
of less than a percent for few-body reaction 
cross-sections~\cite{Gazit1,Gazit2,3H_dis,4He_dis}.
The energy transfer due to elastic scattering is low, $\omega
\sim T^2/m$, and therefore we include only breakup channels
in our calculations.

Since the energy scale of SN neutrinos is much smaller 
than the mass of the $Z$-boson, the neutrino-nucleus interaction 
can be approximated by an effective current-current Hamiltonian. 
The neutrino current is straightforward and results in 
kinematical factors to the cross section. The standard model 
dictates only the formal structure
of the nuclear neutral-current:
\be
J_{\mu}^0 = (1 - 2 \sin^{2} \theta_{W}) \, \frac{\tau_{0}}{2} \,
J^V_{\mu} + \frac{\tau_{0}}{2} \, J^A_{\mu} - \sin^{2} \theta_{W}
\, J^V_{\mu} \,,
\label{eq:NC}
\ee
where the superscripts $A, V$ denote axial and vector currents.
For supernova neutrinos, chiral effective field theory (EFT) of
nucleons and pions offers a consistent approach to nuclear
interactions and electro-weak currents. For historical reasons,
the present calculation uses conventional two- and three-nucleon
interactions and EFT currents, but future applications can be
fully based on chiral EFT. The current approach has
been applied to study electro-weak reactions on $A=2,3,4$ 
nuclei~\cite{Gazit2,PA03}.

We use chiral EFT meson-exchange currents (MEC) at 
next-to-next-to-next-to-leading order. The MEC are based on a 
momentum expansion in $Q/\Lambda$, where $Q \sim 10-20 \mev$ 
is the typical energy in our process of interest, and the 
cutoff $\Lambda$ is of the order of the EFT breakdown scale.
Here, we follow Park {\it et al.}~\cite{PA03} and vary the cutoff over
the range $\Lambda = 400-800 \mev$. In configuration space,
the MEC are obtained from a Fourier transform of propagators 
with a cutoff $\Lambda$. This leads to a cutoff dependence,
which is renormalized by a cutoff-dependent counterterm.
In the present case, all other low-energy coefficients
can be determined from pion-nucleon scattering. The counterterm
$d_r(\Lambda)$ characterizes the strength of a two-nucleon 
contact operator and has been matched to the triton half-life
over this cutoff range. As a check, we reproduce the cutoff 
dependence $d_r(\Lambda)$ of Ref.~\cite{PA03}.

\subsection{Inelastic cross-sections and energy transfer}
\label{cross}

The calculated cross sections are averaged over energy and angle,
assuming a Fermi-Dirac distribution for the neutrinos with zero
chemical potential, temperature $\tnu$, and neutrino momentum $k$,
\be
f(\tnu,k) = \frac{N}{\tnu^3} \frac{k^2}{e^{k/\tnu}+1} \,,
\ee
where $N^{-1}=2 \sum_{n=1}^{\infty} (-1)^{n+1}/n^3$ is a normalization 
factor. The quantities of interest are the temperature-averaged 
cross-sections and energy transfer cross-sections:
\begin{align}
\langle \sigma \rangle_\tnu &= \int_{\omega_{\rm th}}^{\infty} d\omega 
\int dk_i \: f(\tnu,k_i) \, \frac{d\sigma}{dk_f} \,,
\label{eq:cross} \\[2mm]
\langle \omega \sigma \rangle_\tnu &= \int_{\omega_{\rm th}}^{\infty}
d\omega \int dk_i \: f(\tnu,k_i) \: \omega \, \frac{d\sigma}{dk_f} \,,
\label{eq:etcross}
\end{align}
where $k_{i,f}$ are the initial and final neutrino energy,
$\omega = k_i - k_f$ is the energy transfer, and 
$\omega_{\rm th}$ denotes the threshold energy of the breakup reaction.
In Table~\ref{tab:nu_A3}, we present results for the averaged 
neutrino and anti-neutrino neutral-current inclusive inelastic
cross-sections and energy transfer cross-sections as a function of the neutrino
temperature. The difference between the $^3$H and the $^3$He 
cross sections reflects the difference in thresholds between 
the two nuclei. The mirror symmetry between both nuclei is restored 
with higher neutrino energy. The leading contributions to the 
cross section are the axial $E^A_1$, $M^A_1$ and $E^A_2$ multipoles. The 
relative importance of these multipoles varies as a function
of the momentum transfer, and thus as a function of the neutrino 
temperature. In comparison to inelastic excitations of
$^4$He studied in Ref.~\cite{Gazit2}, we find that the cross
sections are about a factor $20$ and $10$ times larger
at temperatures $4 \mev$ and $6 \mev$ respectively (for 
the mean values of $^3$H and $^3$He), and the energy transfer 
cross-sections are $8$ and $2$ times larger.

At low-momentum transfer, the Gamow-Teller operator dominates 
for the cross section, consequently the MEC have a large effect 
of about $16\%$ at a temperature of $1 \mev$. At higher-momentum 
transfer, higher-order multipoles start to play an important 
role. Due to spatial symmetry, the MEC contribution to these 
multipoles is small and the overall effect of MEC decreases 
rapidly to $< 2\%$ for temperatures above $4 \mev$. While not directly 
important here, the asymmetry between the scattering of neutrinos and
anti-neutrinos increases with temperature: The difference in the 
energy transfer grows gradually from $3 \%$ for a neutrino 
temperature of $3 \mev$ to $> 50\%$ for $10 \mev$ temperatures.
Finally, the cutoff dependence of these observables is $< 2\%$ 
for $1 \mev$ and $< 1 \%$ for higher temperatures. This validates
our calculations. We thus estimate the precision of the predicted
cross sections to be a few percent, which also includes estimates of
the numerical accuracy.

\begin{table}
\begin{ruledtabular}
\begin{tabular}{c|cc|cc}
$\tnu\, \text{[MeV]}$ & \multicolumn{2}{c|}{$^3$H} &
\multicolumn{2}{c}{$^3$He} \\[1mm] \hline
1  & 1.97$\times 10^{-6}$ & 1.68$\times 10^{-5}$ 
& 3.49$\times 10^{-6}$ & 2.76$\times 10^{-5}$ \\
2  & 4.62$\times 10^{-4}$ & 4.73$\times 10^{-3}$ 
& 6.15$\times 10^{-4}$ & 5.94$\times 10^{-3}$ \\
3  & 5.53$\times 10^{-3}$ & 6.38$\times 10^{-2}$ 
& 6.77$\times 10^{-3}$ & 7.41$\times 10^{-2}$ \\
4  & 2.68$\times 10^{-2}$ & 3.37$\times 10^{-1}$ 
& 3.14$\times 10^{-2}$ & 3.77$\times 10^{-1}$ \\
5  & 8.48$\times 10^{-2}$ & 1.14                 
& 9.70$\times 10^{-2}$ & 1.25 \\
6  & 2.09$\times 10^{-1}$ & 2.99                 
& 2.35$\times 10^{-1}$ & 3.21 \\
7  & 4.38$\times 10^{-1}$ & 6.61                 
& 4.87$\times 10^{-1}$ & 7.03 \\
8  & 8.20$\times 10^{-1}$ & 13.0                 
& 9.03$\times 10^{-1}$ & 13.7 \\
9  & 1.41                 & 23.4                 
& 1.54                 & 24.6 \\
10 & 2.27                 & 39.3                 
& 2.47                 & 41.2 \\
\end{tabular}
\end{ruledtabular}
\caption{Averaged neutrino- and anti-neutrino-$^3$H and -$^3$He 
neutral-current inclusive inelastic cross-sections per nucleon (A=3), 
$\langle \sigma \rangle_\tnu = \frac{1}{2A} \, \langle \, \sigma_\nu + 
\sigma_{\overline{\nu}} \, \rangle_\tnu$ (left columns), and energy 
transfer cross-sections, $\langle \omega \sigma \rangle_\tnu =
\frac{1}{2A} \, \langle \, \omega \sigma_\nu + \omega
\sigma_{\overline{\nu}} \, \rangle_\tnu$ (right columns), 
as a function of neutrino
temperature $\tnu$, in units of $10^{-42} \, \text{cm}^2$ and $10^{-42} 
\, \text{MeV}\text{cm}^2$ respectively.}
\label{tab:nu_A3}
\end{table}

\subsection{Neutrino energy loss due to inelastic scattering}
\label{dEdx}

\begin{figure*}[t]
\begin{center}
\includegraphics[scale=0.45,clip=]{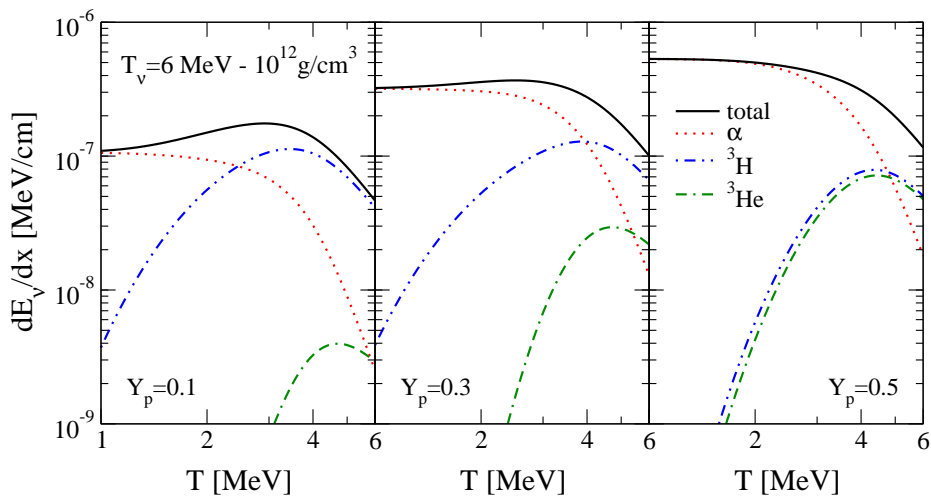}
\caption{(Color online) Neutrino energy loss $dE_\nu/dx$ for inelastic 
excitations of $A=3,4$ nuclei as a function of the matter temperature 
$T$ at a density of $10^{12} \gcq$. We assume that the the neutrino 
energies are characterized by a Fermi-Dirac distribution with a
temperature $\tnu = 6 \mev$. The contributions from $^3$H, $^3$He
and $^4$He nuclei, and the total neutrino energy loss are shown for proton
fractions $Y_p = 0.1, 0.3$ and $0.5$.}
\label{fig:dEdx}
\end{center}
\end{figure*}

We can combine the energy transfer cross-sections with the $A=3,4$
mass fractions of Section~\ref{comp} to calculate the neutrino
energy loss due to inelastic 
excitations of $A=3,4$ nuclei. The neutral-current
cross-sections on $^4$He are taken from Ref.~\cite{Gazit2}, which
is based on the same microscopic input. A neutrino of energy $E_\nu$ 
will lose energy to inelastic excitations, and heat the matter, at 
a rate $dE_\nu/dx$ given by
\be
\frac{dE_\nu}{dx} = n_b \sum_{i={^3}\hspace*{-0.2mm}{\rm{H}},\,
{^3}\hspace*{-0.2mm}{\rm{He}},\,
{^4}\hspace*{-0.2mm}{\rm{He}}} 
x_i \, \langle \omega \sigma \rangle_{i,\,\tnu} \,.
\ee
To explore the effect of mass-three
nuclei on the energy loss, we assume the neutrino energies are 
characterized by a Fermi-Dirac spectrum of temperature $\tnu$,
while the low-density matter may have a lower temperature $T$. 
For simplicity, we neglect the energy transfer from nuclei to 
neutrinos required by detailed balance. 
This is strictly correct only in the limit $T \ll \tnu$.

In Fig.~\ref{fig:dEdx}, the neutrino energy loss due to inelastic
scattering is shown
for a density of $10^{12} \gcq$ and neutrino temperature $\tnu
=6 \mev$, as a function of the matter temperature for various
proton fractions. For $T \gtrsim 4 \mev$, the energy loss
is dominated by the contributions from $^3$H nuclei. The total
abundance of $A=3$ nuclei depends only weakly on the proton fraction
(see Fig.~\ref{fig:composition}), which is reflected in the
weak dependence of the neutrino energy loss as a function of
proton fraction. Finally, for lower densities, mass-three nuclei 
are less abundant (see Fig.~\ref{fig:composition}), and therefore
also their contributions to the neutrino energy loss.

\section{Conclusions}
\label{conclusions}

The virial expansion provides a systematic approach to low-density
nuclear matter in thermal equilibrium, in particular for the
conditions near the neutrinosphere with densities $\rho 
\sim 10^{11-12} \gcq$ and temperatures $T \sim 4 \mev$.
In this paper, we have extended the virial equation of state
of Refs.~\cite{vEOSnuc,vEOSneut} (for matter composed of
neutrons, protons and alpha particles) to include $^3$H and
$^3$He nuclei. We have made model-independent predictions
for the abundance of $^3$H and $^3$He nuclei and predicted
that their mass fractions can be significant near the neutrinosphere.
Our results are directly based on nucleon-$^3$H and nucleon-$^3$He 
scattering phase shifts. In addition, it is interesting that
$^3$H nuclei can be more abundant than free protons at low 
temperatures, since many of the protons are bound in alpha particles 
(and heavy nuclei). In these regions, the rate of electron capture 
may be dominated by capture on $^3$H nuclei~\cite{charged}.

While alpha particles are often more abundant, we have shown that
the loosely-bound $^3$H and $^3$He nuclei dominate the energy transfer 
at low densities through inelastic excitations, and are therefore 
especially important for energy transfer from muon and tau neutrinos.
Our new results for neutrino-$^3$H and -$^3$He neutral-current 
inclusive inelastic cross-sections and energy transfer cross-sections 
are based on microscopic two- and three-nucleon interactions and 
meson-exchange currents, which reproduce the triton half-life.
All breakup channels and full final-state interactions were included
via the LIT method. For temperatures $T \sim 4 \mev$, the predicted 
energy transfer cross-sections on mass-three nuclei are 
approximately one order of magnitude larger compared to inelastic 
excitations of $^4$He nuclei. 

Using the virial abundances and the microscopic energy tranfer
cross-sections, we have found that mass-three nuclei contribute
significantly to the neutrino energy loss due to inelastic
excitations for $T \gtrsim 4 \mev$. Inelastic excitations of 
$^3$H and $^3$He nuclei can be more important than $^4$He 
nuclei at high temperatures, low proton fraction or higher 
densities. To fully assess the role of neutrino breakup of
$A=3$ nuclei, our predicted abundances and neutral-current
cross-sections should be included in SN simulations. The model 
independence of the virial equation of state and the accuracy 
of the predicted cross sections can help to improve the
theoretical nuclear microphysics for SN simulations.

\acknowledgments

We thank Bill Donnelly for useful discussions and the Institute 
for Nuclear Theory, where some of this work was initiated, for 
its hospitality. This work was supported in part by the Natural 
Sciences and Engineering Research Council of Canada (NSERC),
the US Department of Energy under Grant No.~DE-FG02-87ER40365, 
and the ISRAEL SCIENCE FOUNDATION (Grant No.~361/05). TRIUMF 
receives federal funding via a contribution agreement through
the National Research Council of Canada.

\end{document}